\documentclass{article}
\usepackage{authblk}
\usepackage{lscape}
\usepackage{geometry}
\usepackage{needspace}

\title{Speckle Interferometry of 25 Gaia Two-Parameter Potential Binaries}

\author[1]{Paul McCudden}
\author[2]{Russell Genet}
\author[1]{John Major}
\author[3]{Zachary Hartman}
\author[4]{Andy Kovic}
\author[5]{Rick Wasson}
\author[6]{Michael-James Ellis}
\author[2]{Lou Jackson}
\author[1]{Bradley Brungardt}
\author[1]{Zaida Weems}
\author[1]{Astrid Wehlitz}
\author[7]{Evan Wille}
\author[8]{Leon Bewersdorff}
\author[8]{Nick Hardy}
\author[9]{Rachel Freed}
\author[10]{David Rowe}
\author[11]{Thomas C. Smith}
\author[12]{Reed Estrada}
\author[13]{Thomas Meneghini}
\author[13]{Reggie Jones}
\author[13]{Tom Mason}
\author[14]{Dwight Collins}
\author[15]{Mark Copper}

\affil[1]{Colorado Mountain College}
\affil[2]{Gila Community College}
\affil[3]{NASA Ames Research Center}
\affil[4]{Mesa Community College}
\affil[5]{Orange County Astronomers}
\affil[6]{Payson High School}
\affil[7]{University of California, Berkeley}
\affil[8]{OurSky}
\affil[9]{Institute for Student Astronomical Research}
\affil[10]{PlaneWave Instruments}
\affil[11]{Dark Ridge Observatory}
\affil[12]{NASA Neil Amstrong Flight Test Center}
\affil[13]{Mt. Wilson Observatory}
\affil[14]{Presidio Graduate School}
\affil[15]{Magdalena Transit Telescope Project}

\newcommand{\degr}{\ensuremath{^\circ}}

\newcommand{\arcsec}{\ensuremath{''}}

\renewenvironment{abstract}{%
  \needspace{5\baselineskip}%
  \small
  \begin{center}%
    {\bfseries \abstractname\vspace{-.5em}\vspace{0pt}}%
  \end{center}%
  \quotation
}{\endquotation}

\begin{document}
\maketitle
\begin{abstract}
Gaia two-parameter (G2P) stars have cumulative errors in parallax and proper motion so great that only their mean positions were reported in DR3. One potential cause of these high errors is another star as indicated by two intensity peaks in the scans. Speckle interferometry astrometric measurements of 25 G2P stars with high multi-peak percentages were obtained with the 1.5m telescope at Mt. Wilson Observatory. Of the 25 observed G2P stars, seven had no reported Gaia companions within 5.0\arcsec. We found nearby companions for all seven. The 18 other G2P stars had known Gaia companions within 2.0\arcsec. Of these, 13 had separations that agreed closely with the speckle measurements but with some discrepency in position angles, three stars did not agree in either separation or position angle and no companion was detected for the remaining two. Although some of these issues may be resolved in DR4 or DR5, others may be inherent limitations of Gaia capabilities that speckle interferometry observations may be able to fill. 
\end{abstract}
\section{Introduction}\label{intro}
Gaia, the European Space Agency's astrometric telescope, was launched in December 2013, began observations at its L2 station in January 2014, and continued these observations until January of 2025. Analysis of the first 34 months of Gaia operation was made available in June 2024 as data release 3 (DR3) (\cite{fabricius2021},\cite{lindegren2021},\cite{vallenari2023}). DR4 is expected to be released in mid-2026 and will cover 66 months of observation. DR5, to be released in late 2030, will cover the entire 126 months of observation.

As Kareem El-Badry (\cite{elbadry2024}) pointed out in the introduction of his paper, \textit{Gaia's binary star renaissance}, that

\begin{quote}
Binary stars are a cornerstone of stellar mass and radius measurements that underpin modern stellar evolutionary models. \dots They are also ubiquitous, accounting for about half of all stars in the Universe. In the era of gravitational waves, wide-field surveys, and open-source stellar models, binaries are coming back stronger than a nineties trend. Much of the progress in the last decade has been enabled by the Gaia mission, which provides high-precision astrometry for more than a billion stars in the Milky Way.
\end{quote}

Most stars reported in Gaia DR3 have full astrometric solutions that include position, parallax, and proper motions, making them five parameter solutions (G5P stars). G5P stars have revolutionized binary astronomy over the past decade (\cite{elbadry2024}) because these five parallax values allow surveys of stars within a given distance and, thanks to similar proper motions, to establish likely binarity. It is not surprising that most Gaia binary follow-up analysis has concentrated on G5P stars. Below, we detail three examples of such research.

\subsubsection*{Gaia photocenter binaries}
As described by Halbwachs et al. (\cite{halbwachs2023}), by its third data release (DR3), not only had Gaia gathered data on over 1 billion objects, but some 36.5 million stars that were possibly double. These stars were all brighter than G = 19, had a RUWE of greater than 1.4, and were observed over at least 12 visibility periods. RUWE is the renormalized square root of the normalized chi-square of the astrometric fit to the along-scan observations (\cite{lindegren2021}); renormalized such that values close to 1.0 are normal. Of these stars, over half a million have been verified as being short-period binary stars with either full photocenter orbital solutions or significant photocenter orbital curvature over the 34 months of observations by DR3. However, these Gaia DR3 photocenter binaries, essentially by definition, have component separations so small (typically well under 50 mas) that they are well below the resolution limit of Gaia, always imaged as a point source.

\subsubsection*{Wide binaries}
Rather than looking for completed orbits or clear orbital curvature of photocenter binaries in Gaia DR3's 34-month dataset, El-Badry et al. (\cite{elbadry2021}) focused on identifying wider Gaia binaries, compiling an extensive catalog of spatially resolved binary stars within $\approx1$ kpc of the Sun. They used the Gaia catalogue itself to find 1.1 million Gaia G5P eDR3 pairs with a probability of being binary and not mere chance alignments $> 99\%$. These are likely binaries with separations $> 0.4\arcsec$ and, for the most part, $> 1.0\arcsec$. 

\subsubsection*{Re-examining known binaries to discover multiple star systems}
Andrei Tokovinin (\cite{tokovinin2023}) started with a list of 8000 wide binaries within 100 pc with G5P solutions derived from the Gaia Catalog of Nearby Stars. From this he made up a list of multiple star systems where one or both components had large RUWE values, indicators of potential inner subsystem components (Belokurov et al. \cite{belokurov2020}). Tokovinin observed 1243 of these systems with the 4.1-meter Southern Astrophysical Research (SOAR) telescope and discovered 503 new pairs with separations between 0.03\arcsec and 1.0\arcsec, an astonishing triple-system new discovery rate of 40\%.

While most Gaia stars have five-parameter solutions (hence G5P), the cumulative error in the measurements of some of a significant fraction of Gaia stars was so great that the Gaia Data Processing and Analysis Consortium withheld parallax and proper motion measurements from DR3, only providing mean RA and Dec positions, making these Gaia two-parameter (G2P) stars (Chulkov and Malkov \cite{chulkov2022}, Fabricus et al. \cite{fabricius2021}). The most common source of this astrometric measurement error is interference from the light of a nearby on-sky star, often a potential new binary companion. Some stars didn't even achieve G2P status either because they had been insufficiently observed or their position measurements were too variable.

Fabricius et al. (\cite{fabricius2021}) noticed that, compared to Gaia DR2, eDR3 listed many additional sources at small angular separations ($< 0.4\arcsec$). These close neighbors, which were not present in DR2, exceeded the expected distribution from random field star alignments. However, 74\% of these close components at separations $< 0.4\arcsec$ were G2P stars with only two-parameter solutions. Medan and Lépine (\cite{medan2023}) developed a catalog of likely binaries within 200 pc of the sun that included G2P components by using Gaia and 2MASS photometric observations to estimate the missing astrometric parameters in the G2P stars. Medan et al. (\cite{medan2024}) then made speckle interferometry observations of 16 of the potential binaries in their catalog with separations less than 30.0 AU on the 8.1-meter Gemini North and South telescopes. They found, with confidence, that 15 of the 16 were indeed physical binaries.

We have taken yet another approach to studying G2P stars. As described in Section \ref{targets}, we created a list of G2P stars with a high (80-100\%) multi-peak percentage (MP\%). Section \ref{equip} describes the equipment, reduction, and calibration procedures we used to obtain speckle interferometry observations of 25 of the G2P stars on our target list. Section \ref{results} provides our speckle observation results, while Section \ref{discuss} discusses these results, comparing our ground-based astrometry with Gaia's space-based astrometry. Section \ref{conclude} provides our conclusions.

\section{Target Selection}\label{targets}
Stars with G2P solutions were selected from DR3. Furthermore, these stars were selected to have 80 to 100 percent MP\% scans and a magnitude range between 9.5 and 12.9. Stars listed in the Washington Double Star Catalog \cite{hartkopf2001} were excluded for the most part as we were interested in exploring new territory although some were included for comparison. The slow-moving dome of the 1.5-meter telescope at Mt. Wilson Observatory was left parked pointing south to increase observing efficiency, so stars were selected that would transit the dome's slit during our evening observing sessions in June. Of the many potential targets, we observed the 25 G2P stars reported in this paper, as well as 50 known binaries in another research program. 

To avoid repeating the long Gaia DR3 designations, each G2P target was given a number T1, T2, \dots and any nearby Gaia companion ($< 1.0\arcsec$ separation) was given a matching ``C'' companion number. Before assigning these numbers for this paper (but after the observations and analysis), the targets were divided into three logical categories: (I) no matching companion in Gaia, (II) a good match with Gaia in separation, but an increasingly discrepant match in position angle, (III) a poor match with Gaia in both separation and position angle, and (IV) no matching companion detected.

Table~\ref{table:Gaia_data} provides information on the 25 Gaia G2P targets, and the 18 Gaia companions that we extracted from DR3. There were no known close ($<1.0\arcsec$) Gaia companions brighter than 14th magnitude for T1-T5. VPU is the number of visibility periods used where full astrometric solutions are only provided when VPU$>8$ (\cite{lindegren2021}), $5\Sigma$ is the longest semi-major axis of the 5-d error ellipsoid (mas), APS is the number of Gaia parameters (2, 5, or 6), MP\% is the percentage of passes reporting multiple peaks, and $C^*/\sigma_{C^*}$ is the corrected value of the Bp and Rp excess over G normalized by dividing by the standard deviation of $C^*$ (\cite{riello2021}). Values of $C^*/\sigma_{C^*} > 1.65$ are indicators of a non-point source object. A second star in the scan that is unresolved by the Bp and Rp photometry sensors yield very large values of $C^*/\sigma_{C^*}$.
\newgeometry{left=0.25cm,right=0.25cm,top=2cm,bottom=2cm}
\begin{table}
	\caption{Gaia DR3 extracts 25 Gaia G2P targets and 18 Gaia close companions.}
	\label{table:Gaia_data}
	\footnotesize
\begin{tabular}{*{14}{|r}|}\hline
	\multicolumn{7}{|l|}{GAIA TARGET}&\multicolumn{7}{l|}{GAIA COMPANION}\\\hline
	Tgt&	VPU&	$\Sigma5$&	APS&	MP\%&	$C^*/\sigma_{C^*}$&Gaia DR3 Designation&Cmp&VPU&$\Sigma5$&APS&MP\%&$C^*/\sigma_{C^*}$&Gaia DR3 Designation\\
	\hline
	T1&	16&	6.85&	2&	95&	101.2&	4376527797238120320&			&	&&			&&&\\					
	T2&	22&	2.42&	2&	94&	77.8&	4446145815493081472&			&	&&			&&&\\					
	T3-1&	22&	2.13&	2&	85&	45.1&	4446826486207252864&			&	&&			&&&\\					
	T3-2&	15&	1.40&	2&	98&	133.7&	4482139191916699648&		C3-2&  5&     147.80&	2&	100&	153.7&	4482139191925085696\\		
	T4&	29&	5.04&	2&	85&	80.9&	1224087708250606720&			&	&&			&&&\\					
	T5&	15&	2.68&	2&	80&	103.8&	4426488235514632960&			&	&&			&&&\\					
	T7&	16&	2.03&	2&	100&	133.2&	4433728966257750528&		C7&	9&	1.98&	2&	97&	189.5&	4433728966255193856\\	
	T8&	11&	1.38&	2&	98&	162.7&	1164340418192713216&		C8&	11&	1.44&	2&	93&	174.7&	1164340418193799808\\	
	T9&	16&	1.51&	2&	90&	148.1&	1273129324165219968&		C9&	27&	3.58&	2&	78&	104.8&	1273129328460262912\\	
	T10&	14&	2.33&	2&	93&	153.9&	1213718111006872576&		C10&	10&	0.48&	5&	97&	152.3&	1213718111007475584\\	
	T11&	21&	1.91&	2&	93&	114.5&	1203937783282839168&		C11&	6&	1.90&	2&	70&	202.4&	1203937783283326080\\	
	T12&	9&	2.69&	2&	100&	136.1&	6319097322890540928&		C12&	16&	3.24&	2&	96&	140.3&	6319097322891691008\\	
	T13&	18&	4.95&	2&	92&	118.1&	4437793379707330944&		C13&	10&	4.20&	2&	85&	220.6&	4437793379710339840\\	
	T14&	22&	11.06&	2&	94&	158.9&	4571182075942887040&		C14&	9&	0.25&	5&	92&	174.1&	4571182075939811712\\	 
	T15&	21&	2.33&	2&	80&	124.7&	1308686155817944320&		C15&	19&	7.87&	2&	65&	83.3&	1308686160115591040\\	
	T16-1&	22&	1.67&	2&	96&	99.1&	4443360688115386624&		C16-1&	8&	7.19&	2&	59&	237.0&	4443360683820667008\\	
	T16-2&	22&	15.52&	2&	97&	207.2&	4545365985300858112&		C16-2&	20&	19.60&	2&	100&	139.3&	4545365955239147520\\	
	T17&	7&	33.64&	2&	100&	137.3&	6262729965946853248&		C17&	17&	9.6&	2&	58&	51.1&	6262729961645351040\\	
	T18&	6&	1.81&	2&	96&	-----&	4420251668118135808&		C18&	15&	0.52&	5&	94&	114.8&	4420251663823717248\\	
	T19-1&	14&	4.67&	2&	98&	123.3&	1169005199352782848&		C19-1&	7&	230.95&	2&	46&	 -----&	1169005195058227584\\	
	T19-2&	8&	0.24&	2&	100&	-----&	6317521860168221312&		C19-2&	14&	0.05&	5&	98&	74.9&	6317521855872068480\\	
	T20&	29&	2.25&	2&	91&	87.0&	1320858269230075520&		C20&	4&	4376.82&2&	54&	217.4&	1320858269228497408\\	
	T21&	5&	20.68&	2&	99&	-----&	1153076196443974656&		C21&	14&	0.04&	5&	10&	0.5&	1153076196444774144\\
	T22&	18&	24.39&	2&	93&	165.44&	4360470430645686272&		&	&	&	&	&	&	\\
	T23&	21&	14.43&	2&	99&	137.35&	4374763738207810048&		&	&	&	&	&	&	\\
	\hline
\end{tabular}
\end{table}
\restoregeometry

\section{Equipment, Reduction, and Calibration}\label{equip}
Speckle interferometry observations were conducted on June 23, 25, and 26, 2024 at the bent Cassegrain focus of the 1.5-meter telescope at Mt. Wilson Observatory. A ZWO ASI 6200MM Pro CMOS camera (Sony IMX455 sensor), was used with a single Astronomik ProPlanet 642 BP 2850002585 filter with a midpoint transmission of 750 nm.

Each target was imaged 1000 times with 512x512 region-of-interest with exposures typically between 50 and 100ms, along with 300 single-star reference images. Reductions to obtain position angles and separations were made with the Speckle Toolbox (\cite{rowe2015}, \cite{harshaw2017}).

For astrometric calibration, four plate solutions were obtained from four long-exposure images of the periphery of globular cluster M13 or M56 to obtain the plate scale of $0.0306\arcsec/\mathrm{pixel}\pm0.0020$, and camera angle of $175.411\degr\pm 0.073$.

\section{Results}\label{results}
Table~\ref{table:observations} shows both our speckle interferometry results for the target/companion binary pairs, as well as the Gaia values for these same pairs. As mentioned previously, the targets were, after observations and analysis, divided into four logical categories: (I) no matching companion in Gaia, (II) a good match with Gaia in separation, but differences from Gaia measurements in position angle increasing to as much as 12.75\degr, (III) a poor match with Gaia in both separation and position angle, and (IV) companion posited by Gaia but not detected in our data. Note that there are no Gaia values for the seven Category I pairs because there were no known companions brighter than 14th magnitude within $5.0\arcsec$. Two had companions cataloged in the WDS, and thus the remaining five of our observed companions were new discoveries. 

\newgeometry{left=1cm,right=1cm,top=2cm,bottom=2cm}
\begin{table}
	\centering
	\caption{Mt. Wilson and Gaia observations of the 25 targets/companion pairs.}
	\label{table:observations}
	\begin{tabular}{|l|r|r|r|r|r|l|r|r|r|}\hline
	\multicolumn{6}{|l|}{SPECKLE BINARY PAIR OBSERVATIONS}&\multicolumn{4}{l|}{GAIA BINARY OBSERVATIONS}\\\hline

		Target&	Date&	$\rho$ (arcsec)&$\sigma_\rho$&$\theta$ (deg)&$\sigma_\theta$&Cmp&Date&$\rho$ (arcsec)&$\theta$ (deg)\\\hline
		Cat I&&&&&&&&&\\
		T1&	2024.5&0.29&	0.0145&	26.3&3.6173&&&&\\
		T2&	2024.5&0.41&	0.0035&	62.0&0.7239&&&&\\
		T3-1&	2024.5&0.21&	0.0201&	183.2&2.9908&&&&\\
		T4&	2024.5&0.19&	0.0103&	283.9&2.7572&&&&\\
		T5&	2024.5&0.32&	0.0107&	306.5&0.4124&&&&\\
		T22&	2024.5&0.20&	0.0046&	128.7&0.7654&&&&\\
		T23&	2024.5&0.27&	0.0006&	149.5&0.0058&&&&\\
		Cat II&&&&&&&&&\\
		T8&	2024.5&0.53&	0.0066&	162.4&0.1136&		C8&	2016.0&	0.49&	162.20\\
		T3-2&	2024.5&0.30&    0.0025&	264.7&0.2984&		C3-2&	2016.0&	0.33&	265.13\\
		T7&	2024.5&0.51&	0.0111&	52.6&1.6622&		C7&	2016.0&	0.51&	51.91\\
		T10&	2024.5&0.53&	0.0118&	319.5&0.8737&		C10&	2016.0&	0.52&	318.30\\
		T19-2&	2024.5&0.79&	0.0401&	107.8&1.7254&		C19-2&	2016.0&	0.77&	106.54\\
		T12&	2024.5&0.47&    0.0197&	327.7&0.4795&		C12&	2016.0&	0.50&	329.64\\
		T21&	2024.5&1.18&        &	164.6&&		C21&	2016.0&	1.16&	166.59\\
		T9&	2024.5&0.34&	0.0384&	39.6&1.7368&		C9&	2016.0&	0.34&	41.87\\
		T18&	2024.5&0.56&	0.0020&	315.8&1.0806&		C18&	2016.0&	0.55&	311.73\\
		T14&	2024.5&0.40&	0.0006&	265.8&2.0091&		C14&	2016.0&	0.39&	270.50\\
		T11&	2024.5&0.46&	0.0092&	54.7&2.2989&		C11&	2016.0&	0.44&	49.05\\
		T13&	2024.5&0.40&	0.0179&	58.6&2.0667&		C13&	2016.0&	0.36&	67.30\\
		T15&	2024.5&0.23&	0.0087&	151.8&3.9121&		C15&	2016.0&	0.25&	344.55\\
		Cat III&&&&&&&&&\\
		T16-2&	2024.5&0.54&    0.0076&	199.1&0.3544&		C16-2&	2016.0&	0.30&	263.84\\
		T19-1&	2024.5&0.43&	0.0145&	348.2&1.2683&		C19-1&	2016.0&	0.20&	355.42\\
		T20&	2024.5&0.44&	0.0208&	359.1&1.1834&		C20&	2016.0&	0.79&	32.57\\
		Cat IV&&&&&&&&&\\
		T16-1&	2024.5&	&&	&&		C16&	2016.0&	0.50&	246.63\\
		T17&	2024.5&	&&	&&		C17&	2016.0&	0.24&	259.64\\
	\hline
\end{tabular}
\end{table}
\restoregeometry
\section{Discussion}\label{discuss}
	Of the 25 G2P stars we observed, seven (Category I) did not have any known companions in the Gaia catalog within $5.0\arcsec$ brighter than 14th magnitude. Our speckle interferometry observations established that all seven had a close relatively bright companion within $1.0\arcsec$, amounting to a new companion discovery in five of the seven instances (the two others had already been published in the WDS independently of Gaia by other observers). Gaia did list some other nearby faint stars, but they were not within 1.0\arcsec).

This high rate of discovery was not entirely unexpected, as we had selected G2P stars with high multi-peak values ($>80\%$). It seems likely, given their relative brightness, that these newly discovered companions are gravitationally bound. We plan, in future observations, to obtain Sloan multicolor fluxes and, from these, photometric parallaxes which could help establish likely binarity.

It might be noted that Gaia DR4 will feature SEAPipe, the source environment analysis pipeline (\cite{harrison2023}). SEAPipe will provide image reconstruction from transit data. Analysis of these images will find additional sources, and will determine their mean positions, proper motions, parallaxes, and brightnesses . SEAPipe could confirm our discoveries, and more fully characterize many of the current G2P stars, perhaps upgrading them to G5P stars. SEAPipe does not run on all sources however, as a minimal coverage in orientation angle is demanded, so there will still be cases that would benefit from speckle interferometry investigation after the DR4 release.

	Table~\ref{table:differences} compares our successful speckle measurements of the pairs of stars made in June 2024 with Gaia's epoch 2016.0 observations of these same pairs whenever Gaia has published observations of each element and attempts to reconcile discrepancies with rough estimates of intervening motion between the two epochs. Since the stars in each pair have small enough angular separation to suggest they are gravitationally bound, we estimate an upper bound for the possible effect on position angle from gravity as follows. We estimate the temperatures of the stars from Gaia's blue-red light intensity differential. This places the stars on the color-magnitude diagram which provides both an absolute magnitude and an estimate of mass. From the absolute magnitude and the observed magnitude, distance is estimated. We assume our line of sight is perpendicular to the orbital plane when in fact the orbit may be inclined and the actual separation between the stars much greater than appears. Finally we assume that the orbit is circular and thus from Kepler's third law what portion of an orbit might have been traversed in 8.5 years. Again, this maximum possible position angle change (MPA) is a conservative upper bound.

\begin{table}
	\centering
	\caption{Difference between speckle and Gaia binary pair observations.}
	\label{table:differences}
\begin{minipage}{\linewidth}
\centering
\begin{tabular}{|l|*{4}{r|}}\hline
	Target&	$\Delta\rho$ (arcsec)&$\Delta\rho$\%&$\Delta\theta$ (deg)&MPA\footnote{Maximum possible gravitational PA motion} (deg)\\
	\hline
	Category II&&&&\\
	T8&	0.04&8.2\%&0.20&4.46\\
	T3-2&	0.03&9.1\%&0.43&11.80\\
	T7&	0.00&0.0\%&0.69&3.80\\
	T10&	0.01&1.9\%&1.20&2.43\\
	T19-2&	0.02&2.6\%&1.26&0.99\\
	T12&	0.03&6.0\%&1.94&2.48\\
	T21&	0.02&1.7\%&1.99&1.57\\
	T9&	0.00&0.0\%&2.27&3.54\\
	T18&	0.01&1.8\%&4.07&1.69\\
	T14&	0.01&2.6\%&4.70&4.82\\
	T11&	0.02&4.5\%&5.65&8.06\\
	T13&	0.04&11.1\%&8.70&10.86\\
	T15&	0.02&8.0\%&12.75\footnote{Gaia PA computed with respect to brightest component in RP}&36.31\\
	Category III&&&&\\
	T16-2&	0.24&80.0\%&64.74&154.42\\
	T19-1&	0.23&115.0\%&7.22&10.11\\
	T20&	0.35&44.3\%&33.47&2.54\\
	\hline
\end{tabular}
\end{minipage}
\end{table}
Of the 18 G2P stars we observed that had known close companions, 13 (Category II) agreed closely in separation (within 40 milliarcsec) with Gaia DR3 published values. While 8 had position angles within $2.5\degr$ of Gaia's values, the other four disagreed by values ranging from $4\degr$ to nearly $13\degr$. Regressing $\Delta\theta$ on MPA, we obtain an $R^2=0.64$, which suggests that relative motion over 8.5 years between DR3 and our observations may account for a significant portion of the discrepancy between Gaia DR3 measurements and our observations. In this view, it is T18 for which we see unexplained deviation.

We note that the widest discrepancy with Gaia in Category II also has the highest variance among our own measurements. Nevertheless, there may be problems with Gaia's measurements as well.  Tokovinin (\cite{tokovinin2024}) cites two potential causes for such position angle discrepancies. The first is that Gaia may mix the two stars measurements, possibly due to small magnitude differences between the two stars. Our T15 may be a case in point; although both stars are nearly equally bright, Gaia reports that the brighter star in G-band is at the same time fainter in R-band. Tokovinin quotes Holl et al. (\cite{holl2023}) for further explanation. A second contribution to discrepant Gaia position angle estimates advanced by Tokovinin is simply that the Gaia data needs to be interpreted carefully. Many of the stars we observed had low numbers of visibility periods used (VPU), large $\Sigma5$ values in many of the G2P stars, and high RUWE values in the G5P/G6P stars. While some of these issues, such as too few passes, may be resolved in DR4 or DR5, others may be permanent limitations of Gaia capabilities that speckle interferometry observations may be able to fill.

Gaia's reported values differed significantly in both separation and position angle from our measurements of the 3 Category III stars. In two of these cases, T19-1 and T20, Gaia's astrometry is very uncertain by its own estimate with 5-parameter error matrix maximum singular values of 231 and 4377 for their companions when anything larger than 1.2 would have prevented publication of the full 5-parameter astrometry for these stars. That the companions had received only 4 and 7 usable visibility periods may help to explain the elevated singular values. The third star in Category III, T16-2, appears to be quite close with parallax estimated to be 77 mas in \textit{Preliminary Version of the Third Catalogue of Nearby Stars (CNS3)}. Combining this with information from Gaia and WDS suggests an orbit with period roughly in the neighborhood of 80 years and periastron near 2016. The orbital motion of the two stars may then explain Gaia's double digit error matrix singular values as well as the discrepancy with our observation.

Finally there were two stars we observed (T16-1 and T17) for which we were unable to detect a companion, despite the elevated multi-peak percentage, the elevated BP and RP flux excess, C*, and Gaia's publication of companion stars. For T17, our observation faced a dim target, G-magnitude 12.9, moderate delta magnitude of .42 with its companion expected just .24 arseconds away, and a relatively low altitude at time of observation of $39\degr$. Regarding T16, something is wrong somewhere. T19 should be simililar, perhaps even more challenging, but the T16 images are around 15\% brighter and show no sign of multiplicity.

Since our June 2024 observations reported in this paper, we have observed additional G2P stars on the 1.5- and 2.5-meter telescopes at Mt. Wilson Observatory. Instead of observing many high multi-peak percentage G2P stars with close Gaia companions and only a few without known companions, we concentrated entirely on the G2P stars without any known close companions. Also, instead of just observing G2P targets with high multi-peak percentages, we observed some with lower percentages to get a better feel for the relationship between this parameter and the actual multiplicity of G2P stars as well as looking for any relationship between multi-peak percentages and separations detectable by speckle interferometry.

The 2024 speckle interferometry results reported in this paper were based on observations made with a single IR-pass filter. This filter puts observations roughly in line with Gaia's RP bandpass but precludes meaningful photometric analysis. Our own analysis and that of Riello et al. (\cite{riello2021}) have established that for separations less than around 2.0\arcsec, G-band photometry and BP and RP photometry become inconsistant as indicated by $C^*/\sigma_{C^*} > 1$ in Table~\ref{table:Gaia_data}. Below 1.0\arcsec, BP and RP measurements are often completely blended (i.e. equal values for each component). Thus, most of our G2P observations after this paper have been made with Sloan gri filters, allowing a g-i color index and r flux values from our bispectrum analysis. That should allow us to estimate photometric parallax and other parameters (\cite{bailer2018}).

Given the likely errors in position angles and occasional errors in separation for G5P (and G6P) stars with separations $<1.0\arcsec$, as well as the almost totally blended Bp and Rp photometry, it seems likely that even somewhat modest-aperture telescopes equipped for multi-color speckle interferometry can contribute to our understanding of binaries.

\section{Conclusions}\label{conclude}
Seven of the G2P stars we observed had no Gaia companion within 1.0\degr. We discovered, via speckle interferometry, close companions to all seven stars. Two of these stars were known and listed in the WDS. The remaining five may be new discoveries.

The remaining eighteen of the G2P stars we observed did have companions within 2.0\arcsec  listed in the Gaia DR3 catalog. Most of these had separations close to our speckle interferometry measurements but position angle measurements were not uniformly as close. In two cases the published companion was not detected at all. Although some of these issues may be resolved in DR4 or DR5, others may be inherent limitations of Gaia capabilities that speckle interferometry observations may be able to fill.

Our current research is focusing on G2P stars with no known nearby companions. We hope to characterize these newly discovered potential binaries via speckle photometry.

\section*{Acknowledgements}
We are pleased to acknowledge the support of the National Science Foundation (Grant \#2428684) and the Mt. Wilson Institute for the use of their 1.5-meter telescope, kitchen, and dormitory. We thank Magdalena Transit Observatory, Fairborn Institute, Colorado Mountain College, MHA Foundation, and Gravic Inc. for their support of student and instructor travel expenses. We thank Andrei Tokovinin, Robert Buchheim, and others for their helpful suggestions. This work has made use of data from the European Space Agency (ESA) Gaia mission processed by the Gaia Data Processing and Analysis Consortium. This research also made use of the Washington Double Star Catalog maintained at the U.S. Naval Observatory.

\bibliographystyle{plain}
\bibliography{g2p}
\end{document}